\documentclass[10p,final,journal]{IEEEtran}
\usepackage[table,xcdraw]{xcolor}

\usepackage[noadjust]{cite}
\usepackage{subfigure}
\usepackage{epsfig}
\usepackage{physics}
\usepackage{hhline}
\usepackage{amssymb,amsmath,color,graphicx, amsbsy}
\usepackage{algorithm}
\usepackage{algorithmic}
\usepackage{amsfonts}
\usepackage{epstopdf}
\usepackage{multirow}
\usepackage{bbm}
\usepackage{mathtools}
\usepackage{bm}
\usepackage{soul}
\usepackage{amsmath}
\usepackage{comment}
\usepackage[table]{xcolor}

\usepackage{etoolbox}
\usepackage{color}
\usepackage{cite}

\makeatletter

\newcommand{\Rmnum}[1]{\expandafter\@slowromancap\romannumeral #1@}
\makeatother

\newcommand{\minus}{\scalebox{0.75}[1.0]{$-$}}

\DeclareMathOperator{\EX}{\mathbb{E}}
\usepackage{optidef}

\begin{document}

\title{Unsupervised Event Detection,  Clustering, and Use Case Exposition in Micro-PMU Measurements}

\author{Armin Aligholian,~\IEEEmembership{Student Member, IEEE}, Alireza Shahsavari,~\IEEEmembership{Member, IEEE}, Emma Stewart,~\IEEEmembership{Senior Member, IEEE}, Ed Cortez, Hamed Mohsenian-Rad,~\IEEEmembership{Fellow, IEEE}
 \thanks{A. Aligholian, A. Shahsavari and H. Mohsenian-Rad are with
the University of California, Riverside, CA, USA. E. Stewart is with the
Lawrence Livermore National Laboratory, Livermore, CA, USA. E. Cortez is with the Riverside Public Utilities, Riverside, CA, USA. This work is supported in part by UCOP grant LFR-18-548175.
The corresponding author is H. Mohsenian-Rad. 
}

\vspace{-0.8cm}}

\maketitle

\begin{abstract}
Distribution-level phasor measurement units, a.k.a, micro-PMUs, report a large volume of high resolution phasor measurements which constitute a variety of \emph{event signatures} of different phenomena that occur all across power distribution feeders. In order to implement an event-based analysis that has useful applications for the utility operator, one needs to extract these events from a large volume of micro-PMU data. However, due to the \emph{infrequent}, \emph{unscheduled}, and \emph{unknown} nature of the events, it is often a challenge to even figure out what kind of events are out there to capture and scrutinize. In this paper, we seek to address this open problem by developing an \emph{unsupervised}
approach, which requires \emph{minimal prior human knowledge}.
%
First, we develop an unsupervised event detection method based on the concept of Generative Adversarial Networks (GAN). It works by training deep neural networks that learn the characteristics of the normal trends in micro-PMU measurements; and accordingly detect an event when there is any abnormality. We also propose a two-step unsupervised clustering method, based on a novel linear mixed integer programming formulation. It helps us categorize events based on their origin in the first step and their similarity in the second step. The active nature of the proposed clustering method makes it capable of identifying new clusters of events on \emph{an ongoing basis}. The proposed unsupervised event  detection  and  clustering  methods  are applied  to real-world micro-PMU data. Results show that they can outperform the prevalent methods in the literature. These methods also facilitate our further analysis to identify important clusters of events that lead to unmasking several use cases that could be of value to the utility operator.
\vspace{0.25cm}	

\end{abstract}

\section{Introduction} \label{sec:Introduction}

\subsection{Background and Motivation}
\label{subsec: motivation}

Power distribution systems are becoming more active and dynamic due to the increasing penetration of distributed energy sources, electric vehicles, dynamic loads, and etc. This gives rise to various monitoring and control issues. Many of these issues can be addressed by the use of distribution-level phasor measurement units, a.k.a., micro-PMU \cite{hamedreview}.

One of the emerging applications of micro-PMUs is to study ``events'' in power distribution systems. Event-based studies of micro-PMU measurements have a wide range of use cases, such as in situational awareness \cite{situationalawr}, equipment health diagnostics, such as for inverters \cite{irregularity}, capacitor banks \cite{alirezacap}, transformers \cite{transformer}, distribution-level oscillation detection and analysis \cite{Alirezadynamic}, fault analysis and fault location \cite{eventlocation}.

Before one can do any event-based analysis, including for the above use cases in \cite{situationalawr,eventlocation,alirezacap,transformer,Alirezadynamic,irregularity}, one needs to first detect and identify the \emph{events that are of value}. However, this is a challenging task due to at least the following three reasons: 1) most events are \textit{infrequent}; 2) most events are inherently \textit{unscheduled}; and 3) it is often \emph{not known} ahead of time, what kind of events we should seek to find and identify; i.e., we often do not have a prior knowledge about what to look for.

Given the enormous size of measurement data that is generated by micro-PMUs, such as 124,416,000 readings per micro-PMU per day \cite{hamedreview}, the challenges that we listed above call for developing effective data-driven techniques that are 
automated and require minimal prior knowledge.
%
Addressing this open problem is the focus of this paper.

\subsection{Summary of Technical Contributions}
Given the unknown, infrequent and unscheduled nature of events in micro-PMU measurements, in this paper, we propose an inter-connected unsupervised event detection and unsupervised event clustering method for micro-PMU measurements; followed by a comprehensive analysis of the engineering implications of the events in each of the key clusters that we identify from real-world micro-PMU measurements. The main contributions in this paper are listed as follows:

\begin{itemize}

\item 
A novel unsupervised event detection method is developed based on the concept of Generative Adversarial Networks (GAN) by training deep neural networks. Given the infrequent nature of events 
in micro-PMU data, the central idea is to train the GAN models to learn the normal behavior and trends, 
which according to the prior experimental results account for 
99.6\% of the samples 
\color{black} in micro-PMU data. Accordingly, any pattern and signatures that deviates from the captured normal characteristics of the micro-PMU data is marked as an event by the trained discriminator. A set of extracted events by expert knowledge from the real-wolrd data set is used for evaluation. The results show the effectiveness of the proposed event detection method compared to multiple state-of-the-art methods in the literature. The proposed event detection  relies soly on micro-PMU measurements and it does \emph{not} need the network model or prior labeling of the events. \color{black}

\vspace{0.1cm}

\item  A two-step unsupervised clustering method is proposed. In a pre-processing step, events are categorized based on their origin (i.e., the features that are affected by the event), 
 which is obtained from the proposed event detection method. In the second step, in each pre-processed category, a new clustering model is formulated and solved in form of a mixed-integer linear programming (MILP). A rolling based similarity measure, maximum correlation coefficient   (MaxCorr)\color{black}, is used to compare any two events. 
 The proposed clustering method is \emph{active}, i.e., it is capable of identifying new clusters of events on \emph{an ongoing basis}. New clusters are optimally extracted as needed; in order to account for any unknown upcoming events. 
 %
 The experimental results confirm the effectiveness of the proposed clustering model compared to the prevalent clustering methods. The performance
  %
 of the proposed clustering method is evaluated and verified also in comparison with a reference set of clustered events that are obtained by the expert knowledge.

\vspace{0.1cm}

\color{black}
\item The events in each identified cluster are scrutinized in order to unmask their engineering implications and use cases. The origin and the cause of the events are identified to determine their value to the system operator. By implementing the proposed unsupervised approach one can identify the frequency of happening and other statistical characteristics of different event types, extract specific events by combining the event clusters' characteristics and time of occurrence; find rare and unusual events, such as faults and incipient failures and new major loads. It can even identify deficiency in micro-PMU data reporting.

\end{itemize}

\subsection{Literature Review}

The \emph{event detection} component in this paper can be broadly compared with the other data-driven studies such as in \cite{situationalawr,HIFdetection,unequalinterval,rezapartial,arminevent,frquancyevent,vsagcluster,irregularity, detectclass, detectclasslocate, GGL}. Some methods are based on principles in statistics. For example, in \cite{situationalawr}, which we consider as one of the benchmarks for performance comparison in this study, a data-driven statistical event detection method is proposed that is based on absolute deviation around median, combined with dynamic window sizes.  On the other hand, machine learning models, including deep learning models, are getting significant attention in different research areas due to immense increase in the amount of measurement data. Power system is not an exception with massive data collection by measurement units such as smart meters, micro-PMUs and smart inverters, which make researchers capable to implement deep learning model to address issues that are mainly data dependent and complex to solve them with common models. One of the promising applications of deep learning models are anomaly detection which has been implemented in vast scale in smart grid such as, outage detection in the network by using GAN models \cite{outage} and fault detection \cite{ptehrani}, IoT-based occupancy sensor unusual behaviour \cite{IoT} and smart meter anomaly detection \cite{smartmeter}.

\color{black} Some of which are either supervised or semi-supervised. That means, they require either full labeling or partial labeling of the events, e.g., in \cite{HIFdetection,rezapartial}. On the other hand, few event detection methods in the literature that are unsupervised; they are focused on some specific types of events, such as frequency events \cite{frquancyevent}, significant known events such as three phase fault or cap bank switching \cite{detectclasslocate} or cyber attacks \cite{karimipour2019deep}. In contrast, the event detection method in this paper covers a wide range of event types which here, an event is defined rather broadly and may refer to balanced/unbalanced load switching, capacitor bank switching, connection or disconnection of distributed energy resources (DERs), inverter malfunction, a minor fault, a signature for an incipient fault, etc. \cite{emma, PQ, situationalawr}. In \cite{karimipour2019deep}, the authors used symbolic dynamic filters to extract features and dynamic Bayesian networks to learn the system behaviour to detect false data injection. Although the method is unsupervised, this method requires prior knowledge about the dynamics of the system; which is typically not available in practice; such as in the case of the field study and the real-world data analysis in this paper. Similarly, in \cite{detectclasslocate} physical aspect of the system needs to be available in order to detect, classify and localize the abnormalities in the system. Some other unsupervised anomaly detection methods, such as the Generalized Graph Laplacian (GGL) method in \cite{GGL}, are based on determining the graph similarity between the sample windows of the micro-PMU data. We used the method in \cite{GGL} as a benchmark to evaluate the performance of our event detection approach. Other methods that we used as benchmarks include the  unsupervised statistical model in \cite{situationalawr} and the unsupervised learning method introduced in the preliminary conference version of this paper in \cite{arminevent}.

Generative Adversarial Networks (GANs) are broadly studied 
in areas such as image generation \cite{SAGAN}, high-dimensional likelihood-free inference \cite{inf}, medical time-series generation \cite{medical}, and so on.  These models typically focus on the sample generation capability of the GAN model, i.e., the desirable features of the \emph{``generator''} sub-system in the GAN model. However, recent studies have shown that the GAN model can offer other applications also through the desirable features of a \emph{``discriminator''} sub-system. For instance, the GAN model has been used in the recent study in \cite{twodisc} to detect bogus samples, cyber attacks, or general timeseries anomalies \cite{anomalyGAN}. Here, the ultimate goal of the discriminator is to distinguish normal from abnormal samples; Thus, in this paper, the proposed model is focused on carefully adjusting the GAN models for our specific purpose; which is detecting events by discriminator, through learning the normal behaviour of the system.

The \emph{event clustering} component in this paper can be broadly compared with studies such as in \cite{dynamicmonitoring,situationalawr,vsagcluster, faultcluster}. \color{black} In \cite{dynamicmonitoring}, auto encoder-decoder is used for feature extraction; and the latent space of the auto encoder-decoder is used for supervised event classification. In \cite{situationalawr}, supervised support vector classification is used to classify the events based on their source location. In \cite{vsagcluster}, different types of voltage sag events are detected based on a threshold, which is defined by voltage magnitude slope per cycle, then k-means and Ward-method clustering are used to identify the  characteristics of the voltage sag events. The main limitation in \cite{situationalawr,dynamicmonitoring} is that they both require prior event labeling. As for the method in \cite{vsagcluster}, it is focused on voltage sag events. \color{black}In \cite{faultcluster}, the authors used an unsupervised clustering method; however, the focus is on specific faults; such as single-line to ground or line-to-line faults. In contrast, the event clustering method in this paper deals with a wide range of events, including benign yet informative events about the operation status of the power distribution system, its equipment, and its loads. Importantly, no prior knowledge is used in the proposed  method. Therefore, various events can be detected, classified, and characterized; all in  unsupervised fashion. \color{black}In fact, it is designed to explore new events even if they do not match any of the existing clusters through actively searching for new clusters. Thus, the proposed method is well-suited to unmask meaningful use cases based on the outcome of the proposed unsupervised event detection and unsupervised event clustering methods.

\color{black}
This study is fundamentally different from all the previous studies that we have done based on the micro-PMU measurements in our pilot project cite in Riverside, CA. 
The work in \cite{situationalawr}, which is a benchmark for performance comparison in this paper, is about the analysis of events by using \emph{supervised} learning methods. This is in sharp contrast compared to the current paper that is about \emph{unsupervised} learning methods. Unlike in \cite{situationalawr}, where we had to use our field knowledge to do an extensive and time-consuming task of manual event labeling, the method in this paper requires minimal prior knowledge about the power distribution feeder.
The work in \cite{shahsavari2017autopsy, alirezacap,Alirezadynamic} is about scrutinizing some specific types of events, such as capacitor bank switching or certain faults. The assumption in all these three papers is that the events of interests are \emph{already detected and classified}. In this regard, the studies in \cite{shahsavari2017autopsy, alirezacap,Alirezadynamic} can actually benefit from the methodologies that we propose in this paper;    because we can provide each of those use cases with the specific events that we have detected and classified accordingly. 
The work in \cite{eventlocation, farajollahi2017location} is \emph{inherently model-based}; where the focus is one identifying the location of an event that is \emph{already detected and classified}. The location identification methods in these two papers require access to the physical models of the power distribution feeder. In the contrary, the analysis in this paper is \emph{solely data-driven} The work in \cite{farajollahi2018tracking,farajollahi2019linear,akrami2020sparse} is about distribution system state estimation; therefore the analysis is unrelated to this paper; and the results are based on computer simulations, as opposed to the real-world data-driven analysis in this paper. 
Finally, the work in \cite{kamal2020cyberattacks, kamal2020analysis} are about cyber-security attacks against micro-PMUs; which is fundamentally different from the type of studies that we conduct in this paper.

\color{black}
\color{black}

\color{black}

Common event detection schemes, such as moving average filters, statistical change detection methods and PCA, have shown low accuracy due to the verity of event signatures in power distribution feeders; which can be not-perfectly aligned for similar events. Also, methods such as PCA, are meant to represent high-dimensional data with much lower dimensional vectors; which is a suitable approach only if the data lies near a linear manifold in the high dimensional space. However, even with kernel PCA, the non-linearity of the data cannot be modeled appropriately. In general, by increasing the number of training samples, deep learning models usually show higher accuracy compared to the other aforementioned models. Also, GAN models, due to their specific design with a min-max game between a generator and a discriminator, are suitable for learning the essentials of a data distribution. Hence, by a proper problem definition and appropriate generator design, we can articulate real data distribution.  Consequently, the discriminator can distinguish between real data and fake data (not part of the real data distribution). In this regard, using GAN model, alongside with deep learning, is a promising combination for event detection in micro-PMU measurements, which is the approach that we take in this study.
%
\color{black}

Compared to the conference version of this work in \cite{arminevent}, which was \emph{solely about event detection}, this paper has several new  contributions. First, the model architecture and the features are different and result in better performance. Second, to identify the type of detected events, an unsupervised two-step clustering model is proposed. Third, together, the proposed unsupervised event detection and clustering methods enable us to expose use cases and applications of the key event clusters.   

\makeatletter
\renewcommand\subsection{\@startsection{subsection}{3}{\z@}%
{-1.25ex\@plus -1ex \@minus -.2ex}%
{-0 ex \@plus .2ex}%
{\normalfont\normalsize\itshape}}
\makeatletter

\section{Unsupervised Detection Method} \label{detection}

The proposed GAN-based event detection method is developed by training two deep neural networks by using real-world micro-PMU data. In short, the first deep neural network, a.k.a., generator, tries to generate data points that follow the distribution of the real-world data. The second deep neural network, a.k.a., discriminator, tries to distinguish between the generated data points and the real-world data. The architecture and process of the GAN models is as follow:

\vspace{0.05cm}
\subsection{Features} Checking the \emph{magnitude} of voltage and current in micro-PMU measurements is a common option to detect and identify events, e.g., see \cite{situationalawr,jameianom,arminevent}. However, due to the fluctuations in the frequency of the power system, the phase angles of voltage and current are often not used directly. Instead, active power and reactive power are usually used as the two features that involve voltage and current phase angle measurements, besides the magnitude of voltage and current, to detect events in micro-PMU measurements. In this paper, we propose to use power factor as the feature that involves the voltage and current phase angle measurements. Thus, the features across the three phases that we use in this paper are 
\begin{equation}
| V_\phi |, | I_\phi |, \cos(\theta_\phi), \ \ \phi = A, B, C. 
\label{eq:features}
\end{equation}
which denote the voltage magnitude, current magnitude, and power factor in each phase $\phi$, respectively. For notational simplicity, in the rest of paper, we refer to the features in (\ref{eq:features}) for \emph{all the three phases}, without specifying subscript $\phi$. 


\subsection{Generator}
It is a deep neural network that comprises Long Short-Term Memory (LSTM) modules \cite{lstm} as well as dense layers similar to \cite{arminevent}. 
Given a noise vector $z$ from a distribution function $p_z(z)$, such as $z \sim \mathcal{N}(\mu_z , \sigma_z^{2}$), the generator aims to produce samples that follow the distribution of the real-world data. Thus, a neural network $G(z,\theta_g)$ is trained to minimize the following objective function, where  $\theta_g$ denotes the weights of the generator network \cite{GAN}: 
\begin{equation}
        \begin{multlined}
           \frac{1}{N} \sum_{i=1}^{N} \big[log(1-D(G(z_i))) \big].
        \end{multlined}
        \label{gengoal}
\end{equation}
Here, $N$ denotes the number of samples in a batch of training data set. Also, $D$ and $G$ denote the discriminator function and the generator function, respectively. 

\subsection{Discriminator}
It aims to distinguish between the generated samples by the generator and the actual measurements. It contains LSTM modules and dense layers  similar to \cite{arminevent}. Neural network $D(x, \theta_d)$ is trained to report a single value as output. Here, $x$ and $\theta_d$ are the vectors of measurements and the weights of the discriminator network, respectively. The discriminator maximizes the probability of distinguishing the measurement from the data generated by the generator, as: 
\begin{equation}
        \begin{multlined}
             \frac{1}{N} \sum_{i=1}^{N} \big[ log(D(x_i)) + log(1-D(G(z_i))) \big],
        \end{multlined}
        \label{disgoal}
\end{equation}
where {$x_i$ is the $i^{th}$ real sample}. On one hand, the generator tries to minimize (\ref{gengoal}). On the other hand, the discriminator tries to maximize (\ref{disgoal}). Thus, the generator and the discriminator play a \textit{min-max} game over the following function: 
    \begin{equation}
        \begin{multlined}
            V(G, D) = \EX_x\sim p_{data}(x)[log(D(x))] \:  +  \; \: \ \\
             \!\! \EX_x \sim p_{z}(z)[log(1-D(G(z)))].
        \end{multlined}
        \label{GANOF}
    \end{equation}
    
    \vspace{0.05cm}

\subsection{Training and Convergence}
Prior studies have shown that only about 0.04$\%$ of micro-PMU measurements contain events \cite{arminevent}. Therefore, we train all the nine constructed GAN models, one model for each feature, so as to \emph{learn} the characteristics of the \emph{normal} trends in micro-PMU measurements. 
We detect an event when there is an abnormality.
%
For each GAN model, the solution of the min-max game over $V(G,D)$ in (\ref{GANOF}) must satisfy the following two conditions:
\begin{itemize}
    \item \textbf{C1:} For any fixed $G$, the optimal discriminator {$D^*$} is:
    \begin{equation}
        D_G^*(x)=\frac{p_{data}(x)}{p_{data}(x)+p_{g}(x)}.
        \label{global1}
    \end{equation}
    \item \textbf{C2:} There exists a global solution such that: 
    \begin{equation}
        \begin{aligned}
             \min(\mathrel{\mathop{\max_{D}(V(G,D))}}) \Longleftrightarrow p_{g}(x)=p_{data}(x).
        \end{aligned}
        \label{global2}
    \end{equation}
\end{itemize}

\noindent
\color{black}
The training of the GAN model and proofs are explained in details in \cite{GAN}. However, training of GANs is known to be unstable and sensitive to the choices of hyper-parameters, hence, obtaining compelling results such as achieving global optimum and creating a sample distribution close enough to the real data distribution is challenging and requires an assumption that the discriminator is optimal at each step \cite{GANconvergence}. Experimental results in our case with different micro-PMU data set show that local optima and mode collapse situation almost never happen due to non-sharp gradients of the discriminator function around real data points \cite{GANconvergence}.

The choice of the hyper-parameters of the 
GAN model is critical in achieving  an equilibrium. In particular, based on the two criteria in (\ref{global1}) and (\ref{global2}) and  \emph{convergence constant} $\epsilon > 0$ the equilibrium should satisfy the following conditions \cite{GAN}: 
\begin{equation}
\begin{aligned}
& |C(G) - (-log4)| < \epsilon, \\
& |D_g(x) - \frac{1}{2}| < \epsilon.
\end{aligned}
\end{equation}
\color{black}
\subsection{Event Scoring}
Once all the nine GAN models are trained, they provide us with nine distinct event detectors; one per each feature. Each discriminator gives us a \emph{score} as its output, which indicates how close a given window of measurements is to the global optimum that is obtained from (\ref{global1}) and (\ref{global2}). If, for any GAN model, the score is not close enough to the global optimum, then it means that the given window of measurements does \emph{not}  match the normal behavior that is learned by the GAN model; therefore, it 
is deemed to contain an event. \color{black}In this process, the D'Agostino's K-squared test \cite{normtest}, with a significant level of 0.05, is applied to the discriminator output from the training set; and the results show strong evidence of normality. \color{black} Thus, a normal probability distribution function (PDF) is fit to the obtained scores for training set, to have $ \zeta \sim \mathcal{N}(\mu,\sigma^{2})$, where $\mu$ is almost equal to the global optimum and $\sigma$ is small.
\subsection{Algorithm} 
The proposed event detection method is summarized in Algorithm \ref{alg_event_scoring}. The algorithm has two phases. First, a learning phase, in which the GAN models are trained for each feature; and their associated normal PDF are constructed. Second, an event detection phase, in which, for each window $w$ of test data, the scores are calculated by all the nine GAN models and accordingly the \emph{detection vector} is obtained:
\begin{equation}
 \boldsymbol{E}^w_{9 \times 1}=[e^w_{1}, \cdots,e^w_{9}]
    \label{eq:detection vector}
\end{equation}
The detection vector is a $9 \times 1$ \emph{binary} vector, where 9 is the number of features as in (\ref{eq:features}). Entry $e^w_f$ is 1 if an event is detected in $w^{th}$ window and $f^{th}$ feature, otherwise zero. Vector $\boldsymbol{E}_T$ is the set of all detection vectors. It should be noted that, a common choice for $z_p$ in the threshold $\mu \pm z_p \sigma$ is 3, known as the three-sigma rule \cite{threesigma}.

The detection vectors not only show us the existence of event; they also provide us with the inputs that we need for our clustering algorithm; which we will explain in Section \ref{clustering}.
\color{black}
\begin{algorithm}[t]

    \caption{Unsupervised Event Detection}
  \begin{algorithmic}[t]
     \label{alg_event_scoring}
  
      \STATE \textbf{Input:} Training and test data based on the features in  (\ref{eq:features}).
    
      \STATE \textbf{Output:} Event detection vector $E_{9\times1}^w$ for the $w^{th}$ test data.

    \STATE \textbf{// Learning Phase}
    
    \STATE \textbf{Foreach} feature $f$ in (\ref{eq:features}):
    \STATE $ \ \ \ $ Train the $GAN_f$ model
    \STATE $ \ \ \ $ Use discriminator as scoring function $D_{f}^*(\cdot)$. 
    
    \STATE $ \ \ \ $ Calculate the scores for the training data.
    \STATE $ \ \ \ $ Fit a Normal PDF $\mathcal{N}(\mu_f,\sigma_f^{2})$ to the obtained scores. 
    \STATE \textbf{End}
    
    \STATE \textbf{// Detection Phase}
    
    \STATE \textbf{Foreach}  new micro-PMU test data ($w$):

    \STATE $ \ \ \ $ \textbf{Foreach} feature $f$ in (\ref{eq:features}):
    \STATE $ \ \ \ \ \ \ $ Calculate score $s_f^w$ using $D_{f}^*(\cdot)$.
   
    \STATE $ \ \ \ \ \ \ $ \textbf{If} $s_f^w \notin (\mu_f - z_p \delta_f, \mu_f + z_p \delta_f)$ \textbf{Then}
    
    \STATE $ \ \ \ \ \ \ \ \ \ \ \ $ $e_f^w = 1$ // Event
    
    \STATE $ \ \ \ \ \ \ $ \textbf{Else} 
    
    \STATE $ \ \ \ \ \ \ \ \ \ \ \ $ $e_f^w = 0$ // No Event
    
     \STATE $ \ \ \ \ \ \ $ \textbf{End} 
    \STATE $ \ \ \ \ \ \ $ Append $e_f^w$ to $E^w$
    \STATE $ \ \ \ $ \textbf{End} 
    \STATE \textbf{End}
  
  \end{algorithmic}
\end{algorithm}

\color{black}

\color{black}
\subsection{Evaluation metric} 
We use the Matthews correlation coefficient (MCC) \cite{mcc} as the metric to assess accuracy; for both detection and clustering. As explained in \cite{mccbetter}, a common evaluation criteria, such as F1-score, can sometimes be misleading; and MCC is recommended as the more informative accuracy metric in binary classification. On one hand, MCC takes into account the balance ratios of the four elements of the confusion matrix, i.e., true positives (TP), true negatives (TN), false positives (FP), false negatives (FN). On the other hand, for multi-class clustering/classification problems, the general format of MCC is implemented. Therefore, for both event detection and clustering, we use MCC as the evaluation metric: 
\small
\begin{equation}
    MCC=\frac{N_sTr(\psi)-\sum_{k=1}^{K} \sum_{l=1}^{K} \psi_k\psi_l}{\sqrt{N^2_s-\sum_{k=1}^{K} \sum_{l=1}^{K} \psi_k\psi^T_l}\sqrt{N^2_s-\sum_{k=1}^{K} \sum_{l=1}^{K} \psi^T_k\psi_l}},
    \label{eq:mcc}
\end{equation}
\normalsize
\noindent where $N_s$ is the number of samples, $K$ is the number of clusters, $\psi$ is the confusion matrix which is $K \times K$, $\psi_k$ and $\psi_l$ are the $k^{th}$ row and $l^{th}$ column of $\psi$, respectively. It should be mentioned that, for event detection, a special case of general MCC with $K=2$ is used in this study. MCC is a number between \minus 1 and 1; where 1 represents a perfect prediction. MCC is used for the evaluation sets that are extracted by expert knowledge for both event detection and clustering.



\color{black}


\makeatletter
\renewcommand\subsection{\@startsection{subsection}{3}{\z@}%
{1.25ex\@plus 1ex \@minus 1ex}%
{0.8 ex \@plus 0.8ex}%
{\normalfont\normalsize\itshape}}
\makeatletter

\section{Unsupervised Clustering Method} \label{clustering}

Given the detection vectors in Section \ref{detection}, in this section, we develop a two-step event clustering method so that we can later study different types of events in details. 


\subsection{Step \Rmnum{1}: Pre-Processing}\label{step1}

An obvious choice for clustering is to group the events based on their detection vector. For each measurement window $w$ that contains an event, vector $\boldsymbol{E}_{9\times1}^w$ has at least one entry that is one. Accordingly, we can put all the events with the same detection vector in the same category; based on the nine features in (\ref{eq:features}). For example, we put all the events with $\boldsymbol{E}_{9\times1}^w$ $= [1 1 1 \: 0 0 0 \: 0 0 0]$ in the same category because they similarly causes abnormalities only in  voltage magnitude on all phases. 

In theory the detection vector can result in  $2^9-1=511$ possible combinations; when an event is detected. However, based on the physics of the power system; only some of these combinations can actually happen in practice. In fact, our analysis of the real-world micro-PMU data resulted in only a handful of such combinations across thousands of detected events; as we will discuss in details in  Section \ref{subsec: clustering result}. 

Thus, in practice, the above clustering mainly serves as a \emph{pre-processing} in the clustering problem. We often need to further break down a category into several clusters to expose the use case of the events in that category. This is done through a comprehensive clustering optimization in Section \ref{step2}.  

\subsection{Step \Rmnum{2}: Clustering Optimization}\label{step2}

In this section, we explain the similarity measure, the proposed clustering optimization problem formulation, its solution based on exact linearization, the cluster representatives, and the optimum cluster numbers in each category.

\vspace{0.05cm}

\subsubsection{Rolling-Based Similarity Measure}

The key to proper clustering is to accurately measure how similar (or dissimilar) different event signatures are within each pre-processed category. However, this is a challenging task because similar events may not have exact same duration. Events need to be aligned with respect to their shape and their corresponding measurement windows for appropriate similarity assessment.

To address the above two challenges, we propose to first expand the measurement window size for each captured event to make sure that the entire event is included in the measurement window. Once this is done, for each event $i$, we define:  

\begin{equation}
 \boldsymbol{P}_i=
  \begin{bmatrix}
   \alpha_i^{1,1} & \cdots  &\alpha_i^{1,\tau} \\
   \vdots & \ddots & \vdots \\
   \alpha_i^{9,1} & \cdots  &\alpha_i^{9,\tau}
   \end{bmatrix}.
    \label{eq:eventmatrix}
\end{equation}

There are nine rows in $\boldsymbol{P}_i$ corresponding to the nine features in (\ref{eq:features}). The columns correspond to the measurement time instances, where $\tau$ is the maximum expanded window size of the two events that are compared with each other. 

To determine the similarity between two events $i$ and $j$, we need to align matrices $\boldsymbol{P}_i$ and $\boldsymbol{P}_j$, because we do \emph{not} know \emph{where exactly} the event is located within each measurement window. Therefore, we propose to take matrix $\boldsymbol{P}_i$ as fixed, and \emph{roll} matrix $\boldsymbol{P}_j$ in the time axis, one time slot at a time. In other words, in each rolling step, the last column is removed from $\boldsymbol{P}_j$ and appended before the first column in $\boldsymbol{P}_j$; thus, we have $\tau$ rolling steps for each two event comparison.

For each rolling step $k$, where $k = 1, \ldots, \tau$, let us define $c^k_{i,j}$ as the average of the 9 correlation coefficients that can be calculated between each of the 9 rows in $\boldsymbol{P}_i$ and its corresponding row in $\boldsymbol{P}_j$; where  $\boldsymbol{P}_j$ is rolled for $k$ steps. We define \color{black} MaxCorr \color{black} as the rolling-based measure of similarity as: 
\begin{equation}
 MaxCorr_{i,j} = \underset{k = 1, \ldots, \tau} {\text{maximum}} \ c^k_{i,j}; 
    \label{eq:MaxCorr}
\end{equation}
to be used as the similarity measure between events $i$ and $j$.

\vspace{0.05cm}

\subsubsection{Optimization Problem Formulation}

Consider a given category of events based on the pre-processing step in Section \ref{step1}. Suppose there are $I$ detected events in this category and we want to break them down into $C$ clusters. We propose to solve the following clustering optimization problem: 
\begin{mini!}
{u}{\sum_{i=1}^{I}{\sum_{j=1}^{I}{\sum_{c=1}^{C}{u_{i,c}u_{j,c}(1-MaxCorr_{i,j})  }}}}
{}{}
\addConstraint{u_{i,c}}{\in \{0,1\}  }
\addConstraint{\quad \quad \sum_{c=1}^{C}{u_{i,c}}}{=1 }{\quad \forall i}. 
\end{mini!}
where $u_{i,c}$ is a binary variable. It is one if  event $i$ is in cluster $c$; otherwise it is zero. Problem (12) minimizes the sum of the distances between the events, measured as 1-$MaxCorr_{i,j}$,  across different clusters. The constraint in (12c) assures that each event is assigned to only one cluster. Problem (12) is a MINLP. 

\subsubsection{Exact Linearization}
To enhance computational performance, the MINLP in (12) is replaced with an \emph{exact equivalent} Mixed Integer Linear Programming (MILP), as follows: 
\begin{mini!}
{u,t}{\sum_{i=1}^{I}{\sum_{j=1}^{I}{\sum_{c=1}^{C}{t_{i,j,c}(1-MaxCorr_{i,j})}}}  \label{eq:optlpOF}}
{}{}
 \addConstraint{u_{i,c}, t_{i,j,c}}{\in \{0,1\}  }
 \addConstraint{\sum_{c=1}^{C}{u_{i,c}}}{=1  }{\quad \forall   i }
\addConstraint{u_{i,c}+u_{j,c}-t_{i,j,c}}{\leq 1 }{ \quad \forall  i,j,c }
\addConstraint{-u_{i,c}-u_{j,c}+2t_{i,j,c}}{\leq 0  }{\quad \forall  i,j,c. }
\end{mini!}
where the nonlinear product of $u_{i,c}$ and $u_{j,c}$ in the objective function is replace with linear term $t_{i,j,c}$. The linear constraints in (13d) and (13e) are used to make sure that $t_{i,j,c}$ is indeed equal to such product in order to assure an \emph{exact} linearization. Problem (13) can be solved using any MILP solver for a set of detected events in a given time period as training set.

\vspace{0.05cm}

\subsubsection{Cluster Representatives} Once the clusters are obtained by using the training data and solving the MILP problem in (13), we define a representative for each cluster  to speed up the process of clustering incoming events. Thus, the new events are compared to a few cluster representatives rather than to all events through (13). 
To determine the optimum representative for each cluster, we solve the following optimization problem:
\vspace{-0.4cm}
\begin{mini!}
{v}{\sum_{i=1}^{I}{\sum_{j=1}^{I}{\sum_{c=1}^{C}{u_{i,c}v_{j,c}(1-MaxCorr_{i,j})}}}  \label{eq:optlpOFwithcand}}
{}{}
\addConstraint{v_{i,c}}{\in \{0,1\} }
\addConstraint{\sum_{i=1}^{I}{v_{i,c}}}{=1 \quad \forall   c }
\end{mini!}
%
%
%
Variable $v_{j,c}$ is binary. It is one, if event $j$ is the representative event for cluster $c$, and zero otherwise. Constraint (14c) is used to make sure that there is only one representative for each cluster. 
Notice that $u_{i,c}$ is parameter, not a variable, in this optimization problem; because the clusters are already formed. Therefore, problem (14) is an MILP by construction. 

\vspace{0.05cm}

\subsubsection{Number of Clusters ($N_c$)} So far, we have assumed that the number of clusters, i.e., parameter $c$ is fixed. However, we do obtain the optimal number of clusters in our proposed method. This is done by  solving the optimization problem in (13) with respect to different number of clusters. Then, the optimal number of clusters is determined based on the silhouette values of the clusters. Subsequently, cluster representatives is identified for the optimally obtained cluster by using (14). 

\vspace{0.1cm}

\subsection{Active Clustering}
\color{black}Since the proposed event clustering method is active, it can create new clusters. Given the large number of events that are detected in micro-PMU measurements for our dataset, it is computationally prohibitive to cluster all of 15 days events at the same time. On the other hand, by training a subset of the detected events and assign the new events to the trained cluster, those new types of events would be assigned to a wrong cluster. To address these issues, we solved the clustering optimization problem only on the first day in our dataset to set up a base for the event clusters. The newly detected events would be compared to each base cluster representatives and they will be assigned to the closest cluster. Unless, if \color{black} MaxCorr \color{black}s of a new event is less than a threshold ($\varphi$) for every existing cluster representative, then a new cluster is created. In practice, such new cluster is added only occasionally, which shows the common events are almost appear in every day. However, the newly added clusters are usually those weakly or rare events. Furthermore, the clusters can be updated using the complete optimization-based approach periodically once every few days in order to pick the optimal representative
s for each cluster.

\color{black}
\begin{algorithm}[t]
\color{black}

    \caption{ Unsupervised Event Clustering}
  \begin{algorithmic}[t]
     \label{alg_active_clustering}
  
      \STATE \textbf{Input:} Event detection vectors $\boldsymbol{E}_T$ from (\ref{eq:detection vector}) \\ \ \ \ \ \ \ \ \; Event time-series data,\\ \ \ \ \ \ \ \ \; Number of clusters \\ \ \ \ \ \ \ \ \; Similarity threshold ($\varphi$).
    
      \STATE \textbf{Output:} Clusters and their representatives \\ \ \ \ \ \ \ \ \ \ \ \ Silhouette value.

    \STATE \textbf{// Learning Phase (Offline)}
    \STATE Create categories based on unique sets in $\boldsymbol{E_T}$.
    \STATE Assign each event to its category.

    \STATE \textbf{Foreach} $n$ from 1 to $N_c$:
    \STATE $ \ \ \ $\textbf{Foreach} category in $\boldsymbol{E_T}$:
    
    \STATE $ \ \ \  \ \ \ $ Cluster the events based on (13).
    \STATE $ \ \ \  \ \ \ $ Determine the cluster representative based on (14).
    \STATE  $ \ \ \ $\textbf{End}
    \STATE \textbf{End}
    \STATE Calculate Silhouette value for all possible combinations. 
    \STATE 
    Set the number of clusters and their  representatives 
    
    \STATE \textbf{//Active clustering Phase (Online)}
    
    \STATE \textbf{Foreach}  new event  in test data:

    \STATE $ \ \ \ $ \textbf{If} the detection vector of new event is in $\boldsymbol{E_T}$ \textbf{Then}
    \STATE $ \ \ \ \ \ \ $ Calculate MaxCorr with all representatives.
    \STATE $ \ \ \ \ \ \ $ \textbf{If} all calculated MaxCorrs are less than $\varphi$ \textbf{Then}
    \STATE $ \ \ \ \ \ \ \ \ \ $Make a new cluster in the related category.
   \STATE $ \ \ \ \ \ \ \ \ \ $Set new event as the cluster representative.
   \STATE $ \ \ \ \ \ \ $ \textbf{Else}:
   \STATE $ \ \ \ \ \ \ \ \ \ $ Assign new event to the closet cluster.
   \STATE $ \ \ \ \ \ \ $ \textbf{End}
   \STATE $ \ \ \ $ \textbf{Else} :
   \STATE $ \ \ \ \ \ \ $ Create a new category.
   \STATE $ \ \ \ \ \ \ $ Create a cluster with new event as representative
   \STATE $ \ \ \ $ \textbf{End} 
   \STATE \textbf{End}
  \end{algorithmic}
\end{algorithm}

\color{black}

\color{black}
\section{Experimental Results} 
\label{Results}

The proposed event detection and clustering methods are applied to 1.2 billion measurements over 15 days of real-world micro-PMU data. Fourteen days of data are used for training the event detection method and one day of data is used to test it. One day of data is used for cluster optimization; and active clustering is done for the rest of the data.

\subsection{Parameters Detail} \label{sec: Parameters Detail}
\color{black}
The architecture of the GAN model has two parts. The generator starts with a dense layer of size 40, three layers of LSTM with 32, 64 and 128 modules, and a dense layer of size 256. The discriminator is in reverse order; the only difference is that the last layer in the discriminator is a dense layer with size 1. All activation functions are LeakyReLU except the last layer in the discriminator; which is sigmoid. In the LeakyReLU functions, the slope of the leak is set to 0.2 in all models. For tuning the hyper-parameters, we used the \emph{coarse-to-fine} method. In this method, we first randomly chose a set of values for each parameter. Then we narrowed down the choices to a smaller subset based on the obtained results. This procedure was repeated until we achieved the desired value for each parameter. It should be mentioned that, depending on the hyper-parameter, scaling can be helpful, such as log scale for learning rate. This can help fasten the search for suitable values of choice. 
%
%
The learning rate $\alpha$ is set to 0.0002 for Adam optimizer and $\beta_1$ is set to 0.5 for better stability in training. A critical parameter when it comes to capturing the events appropriately is window size, which first it should be wide enough to capture the essential signatures of an event and second it should be small enough to prevent event synchronicity and high computational time. By analyzing different window sizes for different set of micro-PMU data, the best result in terms of accuracy and reasonable computational time, is 40 data points. Also, in order to assure that events are not overlooked, we consider that each window has 20 data points overlap with the previous window. All GAN models are developed with Tensorflow in Python by using Nvidia GTX 1050 ti GPU and core i-7 CPU.

\subsection{Event Detection Results} \label{subsec: detection result}
Table \ref{table:detection} shows the MCC for the proposed event detection method, in comparison with the benchmark methods in \cite{situationalawr,arminevent, GGL}. A total of 1200 reference events are visually extracted by expert knowledge within 6 hours (64800 window samples of 40 timeslots) to evaluate the performance of event detection. The proposed method outperforms the methods in \cite{situationalawr}, \cite{arminevent} and \cite{GGL}.
The combined training time of all 9 GAN models is 2 hours. Once the initial training is done, it takes less than 4 milliseconds to determine a new incoming sample as normal or event; i.e., the detection time is 4 milliseconds. The detection time for \cite{situationalawr} and \cite{arminevent} is 3 and 10 milliseconds, respectively. Thus, the proposed method maintains the same level of computational complexity; but it achieves much better accuracy.
It should be added that, to have a fair comparison with the model in \cite{GGL}, we analyzed different window sizes for the aforementioned method; and as we increase the sample numbers, the accuracy is improved, however, the \emph{rate} of the improvement in accuracy was decreasing, in other words, accuracy does not change significantly by extending the window at a certain point; Also, the training time is ascending as well. Thus, the best performance with the same training time as GAN models are considered for the method in \cite{GGL}. Another point about the method in \cite{GGL} is that, due to the use of similarity graph, the method in \cite{GGL} needs to be re-trained every time which makes it impractical for detecting events with new upcoming data in an online mode, however, for offline mode this method has the lowest False Positive (FP).

\color{black}
An interesting observation when we compare the proposed event detection model with the model in \cite{arminevent} is that, the choice of the independent features in (1), in particular the use of $\cos(\theta)$ instead of active power and reactive power, improves the accuracy of event detection. It also improves the independence in the outputs of the trained GAN models. This makes the resulting detection vectors to even enhance the performance of the subsequent clustering method. \color{black} One of the main advantages of the proposed model compared to the GGL method in \cite{GGL} is the aspect of learning  the normal operation of the system. Although the GGL method has a very low false positive rate, the number of its  true positives is lower than the method in \cite{arminevent}. The number of its true positives is also lower than the proposed method in this paper. The reason is that, when several events happen continuously, i.e., they happen back to back, such as the events in Fig. 12, the GGL method would train its similarity matrix based on the considered window sample. In this case most of the samples are events, thus, events are \emph{not} anomaly anymore for the GGL method. As a result, there would be higher score value of similarity among such event-containing back-to-back window samples. This leads to a lower true positive rate for the GGL method. On the contrary, the proposed model in this paper considers each sample \emph{individually} and it compares each sample with learned signature of the normal samples by the GAN models. This improves true positive rate. It should be noted that, the statistical method in \cite{situationalawr} has reasonable accuracy in detecting most of the three phase events that are \emph{balanced}; due to the fact that the method in \cite{situationalawr}  was not designed to particularly detect unbalanced events. The method in \cite{situationalawr}  performs poorly also to detect events with low magnitude. Both of these issues are resolved in this paper. The proposed method in this paper is particularly well-positioned to detect unbalanced events as well as high frequency but low magnitude events. Given the fact that it is common to have unbalanced events in power distribution systems, this particular advantage of the proposed method is of importance in real-world applications.

\color{black}
\begin{table}
\centering
\caption{\color{black} Event Detection Information for Confusion Matrix, \\ Precision, Recall and MCC}
\label{table:detection}

\scalebox{0.7}{
\renewcommand{\arraystretch}{1.5}
\small
\color{black}
\begin{tabular}{|c|c|c|c|c|}

\hline  Metric & Statistical \cite{situationalawr} & GGL \cite{GGL} & Enhanced Method \cite{arminevent} & Proposed Method\\
\hline TP & 311 & 990 & 1033 & 1132\\
\hline FN & 889 & 210 & 167 & 68\\
\hline FP & 210 & 1 & 56 & 36\\
\hline TN  & 63390 & 63599 & 63544 & 63546\\
\hline Precision & 0.596 & 0.998 & 0.948 & 0.947\\
\hline Recall & 0.259 & 0.825 & 0.861 & 0.943\\
\hline MCC & 0.386 & 0.906 & 0.901 & 0.955\\
\hline
\end{tabular}}
\renewcommand{\arraystretch}{1}
\end{table}
\color{black}

\vspace{0.2cm}

\subsection{Event Clustering Results} \label{subsec: clustering result}

The proposed event clustering method is applied to the captured events in Section \ref{subsec: detection result}; and its performance is compared with the following prevalent clustering methods in the literature: kNN \cite{kmeans}, k-Medoids \cite{kmed}, and fuzzy-k-Medoids \cite{fuzzyclustering}. Different similarity measures are also considered: euclidean, DTW \cite{dtw}, soft-DTW \cite{softdtw}, and \color{black} MaxCorr. In order to compare clustering results, different indices are implemented in the literature such as Jaccard Index, Adjusted Rand Index, Fowlkes Mallows Index, Normalized Mutual Information and Silhouette index; where  the last one, i.e., the silhouette index is generally known to show better results within variety of data sets \cite{clusteringcriteria}. However, if a labeled evaluation set is available, the analysis and assessment of the clustering model is more intuitive and informative. Thus, in this paper the comparison is conducted over 4000 reference events that are visually clustered with expert knowledge. These events are clustered after being detected by the proposed event detection method.\color{black}

Table \ref{table:clustering} shows the MCC for different clustering methods. Two observations can be made based on the results in this table. First, the clustering methods are almost always more accurate when \color{black} MaxCorr \color{black} is used as  similarity measure. Second, our proposed clustering method outperforms kNN, k-Medoids, and fuzzy-k-Medoids for any 
similarity measure. \color{black} The computational time to \emph{train} the KNN, k-medoids, Fuzzy k-medoids, and the proposed models (when MaxCorr is considered as similarity measure) are 5 minutes, 7 minutes, 15 minutes, and 65 minutes, respectively. Note that, training is done \emph{offline}. 
Therefore, the higher 
accuracy of the proposed model does \emph{not} cause higher computational time during the operation. Importantly, recall that the proposed clustering method is \emph{active}. In fact, when it comes to  clustering new upcoming events that require creating new clusters, which is done \emph{online} and during operation, all of the above methods have almost the same computational time; which are less than 4 milliseconds. \color{black}
\begin{table}
\centering
\caption{\color{black} \small MCC in Event Clustering for Different Methods and Different Distance Criteria}
\label{table:clustering}
\scalebox{0.7}{
\renewcommand{\arraystretch}{1.5}
\normalsize
\color{black}
\begin{tabular}{|l|c|c|c|c|}

\hline  Distance & KNN  & k-medoids  & Fuzzy k-medoids & Proposed Method\\
\hline Euclidean & 0.451 & 0.543 & 0.522 & 0.447\\
\hline DTW  & 0.584 & 0.863 & 0.871 & 0.911\\
\hline soft-DTW  & 0.579 & 0.861 & 0.871 & 0.888\\
\hline MaxCorr & 0.645 & 0.887 & 0.882 & 0.938\\
\hline

\end{tabular}}

\vspace{0.05cm}
\end{table}

\subsection{Analysis of Identified Clusters}

A total of nine detection vectors were observed among all the events which they are denoted by $\mathbf{E}_1$ to $\mathbf{E}_9$, as shown in Table \ref{table:clustercategories}. As part of the pre-processing step in Section \ref{step1}, these detection vectors result in five categories, denoted by Category \Rmnum{1} to Category \Rmnum{5}, as shown  on the last column in Table \ref{table:clustercategories}. Categories \Rmnum{1}, \Rmnum{2}, and \Rmnum{3} include \emph{balanced} events; while Categories \Rmnum{4} and \Rmnum{5} include \emph{unbalanced} events. 

The optimization-based clustering in Section \ref{step2} is then applied to the above five categories. It resulted in identifying a total of \textbf{16 final clusters}. In this regard, Category I is divided into six clusters; Category II is divided into three clusters; Category III is one cluster by itself; Category IV is divided into three clusters; and Category V is divided into three clusters.  

Next, we use the above clustering results to scrutinize and expose the use cases for the events within each cluster.

\begin{table}
\centering
\caption{Cluster Categories from Pre-Processing}
\label{table:clustercategories}
\scalebox{0.8}{
\begin{tabular}{|c|c c c|c|c|}

\hline   
Detection 
&  & Features &  & Number  & Pre-Processing \\

Vector  & $\abs{V}$ & $\abs{I}$ & $cos(\theta)$ & of Events & Category\\

\hline  $\boldsymbol{E}_1$  & [111 & 111 & 111] & 34242 & \Rmnum{1} \\
\hline $\boldsymbol{E}_2$ &[111 & 000 & 000] & 12270 & \Rmnum{2}\\
\hline  $\boldsymbol{E}_3$ &[000 & 111 & 111] & 809 & \Rmnum{3}\\
\hline  $\boldsymbol{E}_4$  &[000 & 100 & 100] & &\\
    $\boldsymbol{E}_5$  &[000 & 110 & 110] & 13956 & \Rmnum{4}\\
   $\boldsymbol{E}_6$  &[000 & 011 & 011] & &\\
  
\hline   $\boldsymbol{E}_7$  &[000 & 000 & 111] & &\\
  $\boldsymbol{E}_8$  &[000 & 000 & 110] & 52 & \Rmnum{5}\\
  $\boldsymbol{E}_9$  &[000 & 000 & 011] & &\\
 \hline
\end{tabular}}
\end{table}

\subsection{Use Case Exposition: Six Clusters in Category \Rmnum{1}}

Six clusters are identified in Category \Rmnum{1}; denoted by Clusters \#1 to \#6. Clusters \#1 and \#2 can help identify different load types. Clusters \#3 and \#4 can reveal malfunctions in the operation of capacitors. Cluster \#5 can help identify a specific two-step transient events. Cluster \#6 can identify oscillations. 

\subsubsection{Identifying Different Load Types}
\begin{figure}[t]
\begin{center}
\includegraphics[scale=0.54]{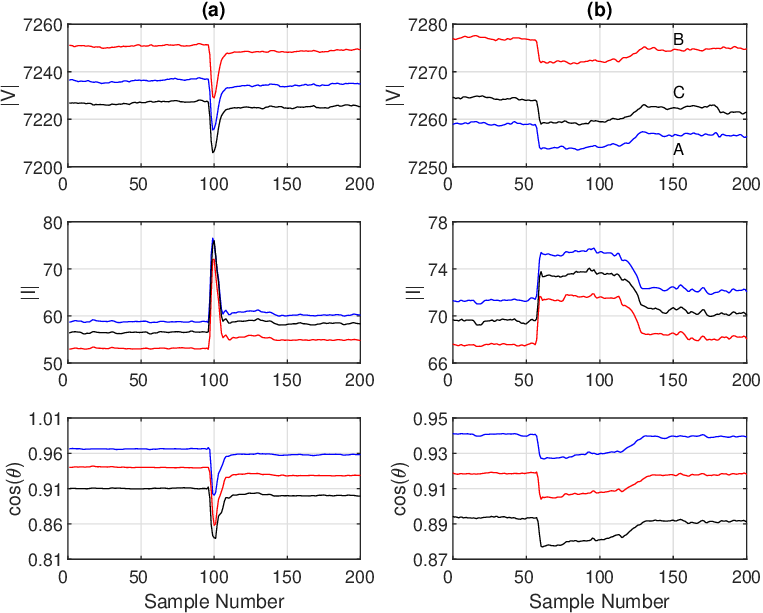}
\end{center}
\vspace{-0.25cm}
\caption{Examples of load switching events: (a) inrush current in Cluster \#1; (b) long transient with a plateau in Cluster \#2.} \label{fig:inrmed}
\vspace{-0.3cm}
\end{figure}

\begin{figure}[t]
\vspace{-.2cm} \centering
\hspace*{-.3cm}
\includegraphics[width=0.95 \columnwidth]{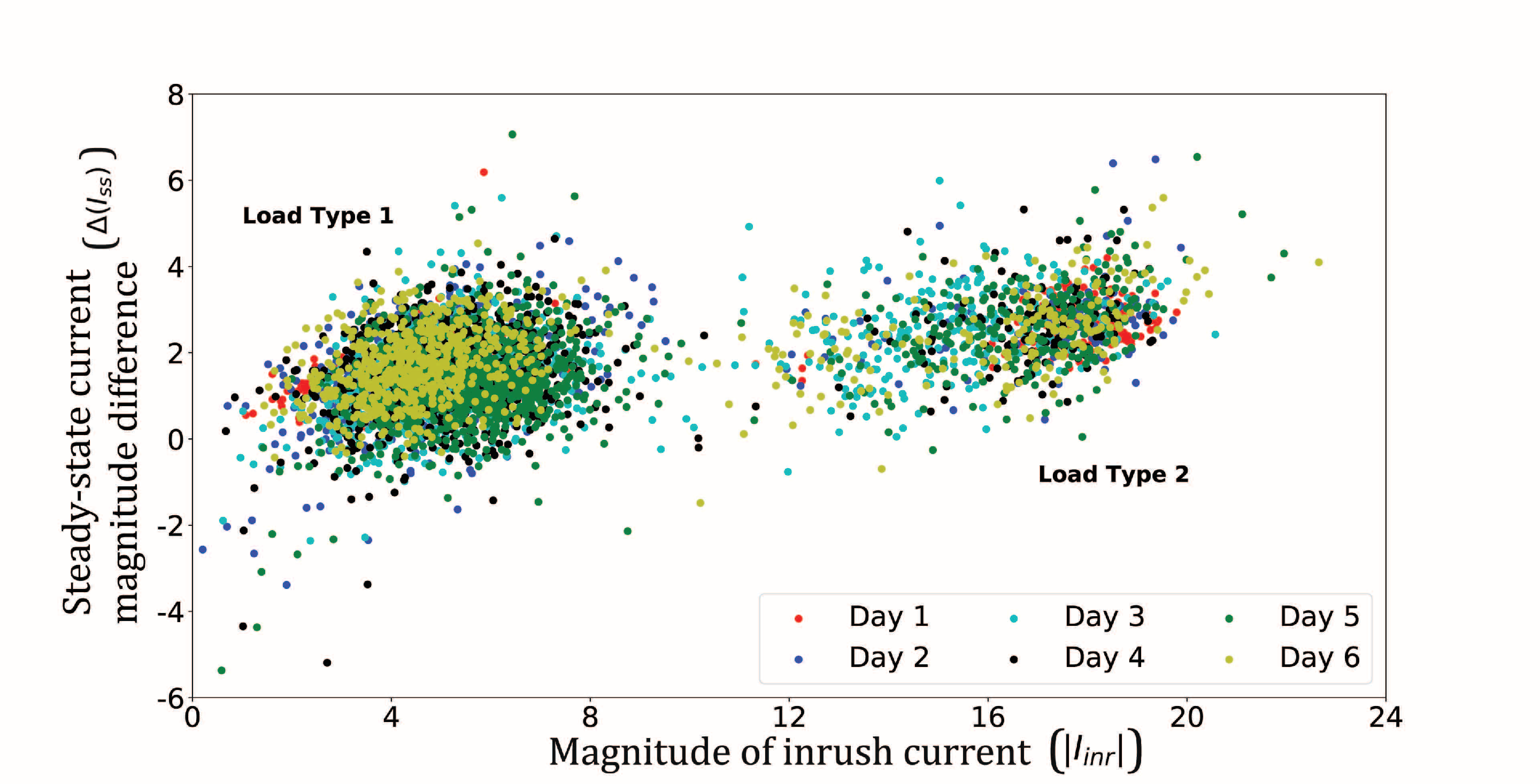}
  \vspace{-0.15cm}
  \caption{Identifying two major load types based on Cluster \#1.}
   \label{fig:inrstat}
   \vspace{-0.15cm}
\end{figure}

Fig. \ref{fig:inrmed}(a) shows an example for Cluster \#1, which is the most frequent event in this system. It is the inrush current from load switching. The transient time of these events is less than 10 time slots, i.e., $83.3$ msec, and one pinnacle which illustrates the magnitude of inrush current. Fig. \ref{fig:inrstat} shows the scatter plot for the change in the steady-state current, i.e., $\Delta(I_{ss})$, versus the magnitude of inrush current, i.e., $\abs{I_{inr}}$ during 6 different days. As it can be seen, Cluster \#1 can it self be divided into two main sub-clusters which show two major types of loads in this cluster.

\begin{figure}[t]
\vspace{-.3cm} \centering
\hspace*{-.2cm}
  \includegraphics[width=0.95 \columnwidth]{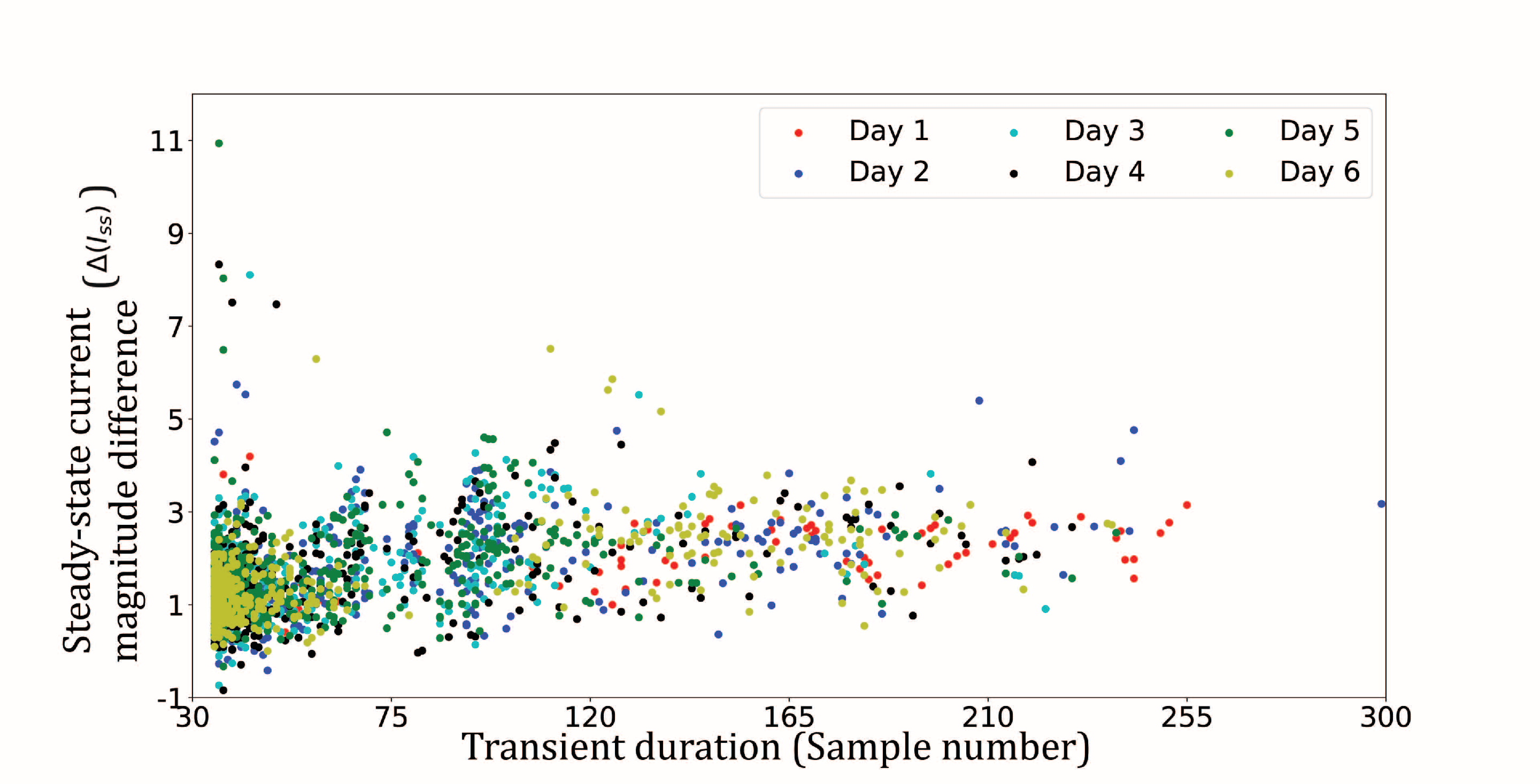}
  \vspace{-0.15cm}
  \caption{Scatter plot for the events in Cluster \#2 over 6 days.}
   \label{fig:medstat}
   \vspace{-0.25cm}
\end{figure}

Fig. \ref{fig:inrmed}(b) shows an example for Cluster \#2. It is for the load types that create much longer transient period to switch and creates a \emph{plateau}; which is very different from the inrush current in Cluster \#1 with a \emph{pinnacle}. 
Fig. \ref{fig:medstat} shows a scatter plot for the events in Cluster \#2. On the y-axis it shows the change in steady-state current, \emph{before} and \emph{after} the event, which is denoted by $\Delta(I_{ss})$. The x-axis is the length of the transient period of the event. There is a dense concentration area, where $\Delta(I_{ss})$ fluctuates at around 1.5 A. This observation empowers the system operator to more readily detect any abnormalities in this cluster, with regard to $\Delta(I_{ss})$ and transient duration, such as multiple simultaneous load switching.

\subsubsection{Capacitor Bank State of Health Monitoring}

\begin{figure}[t] 
\begin{center}
\includegraphics[scale=0.54]{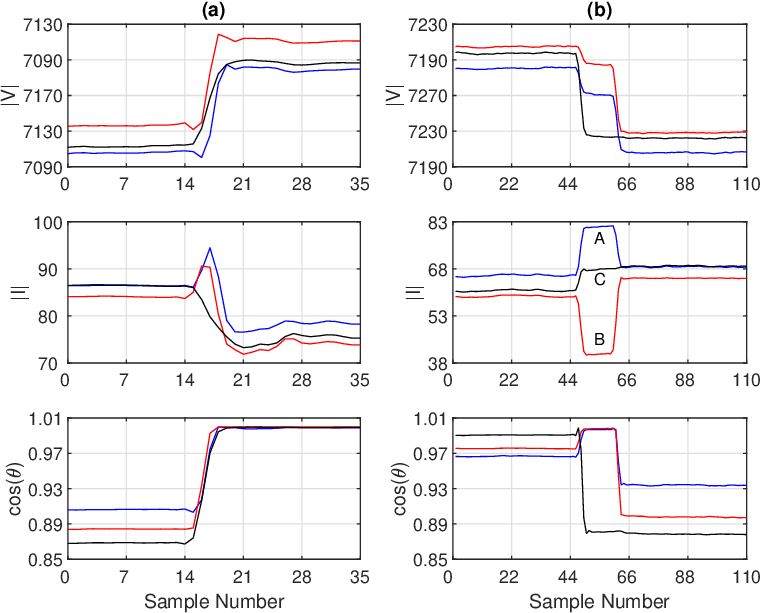}
\end{center}
\vspace{-0.25cm}
\caption{Monitoring the operation and health of a capacitor bank based on Clusters \#3 and \#4: (a) switch on; (b) switch off.} 
\label{fig:capbank}
\end{figure}

Figs. \ref{fig:capbank}(a) and (b) show examples of clusters \#3 and \#4, which are related to capacitor bank switching `on' and switching `off' events, respectively. Capacitor bank switching occurs on a daily basis.  

Monitoring the switching actions of capacitors can not only keep the utility operator informed of switching status of the capacitor banks; it can also help to evaluate their state of health. For example, consider the capacitor bank switching off event in Fig. \ref{fig:capbank}(b). We can see that there is a relatively long \emph{overshoot} on Phase A current  and a relatively long \emph{undershoot} on Phase B current  before the capacitor is  de-energized. This is likely due to a \emph{malfunction} in the switching control mechanism at the capacitor bank, c.f. \cite{alirezacap}. By clustering all the capacitor switching events, we can conduct statistical analysis on the characteristics of such transient switching responses and dispatch the field crew to examine the capacitor bank switching controller and perform repairs.

\subsubsection{Two-step Events}
\begin{figure}[t] 
\begin{center}
\includegraphics[scale=0.54]{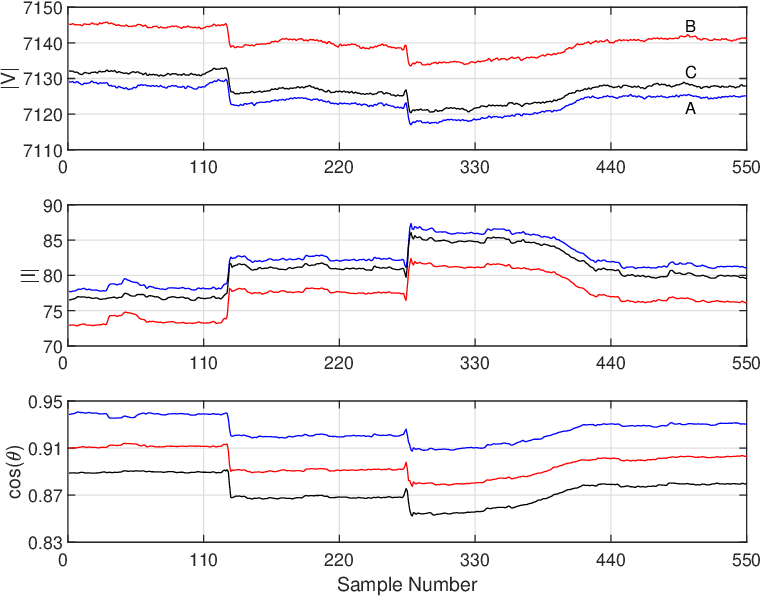}
\end{center}
\vspace{-0.25cm}
\caption{An example for the two-step event in Cluster \#5.}
\vspace{-0.1cm}
\label{fig:twostepmed}
\end{figure}

Fig. \ref{fig:twostepmed} shows an example of the special load in Cluster \#5. This special type of load has two separate but subsequent steps. By using the proposed unsupervised event detection and unsupervised event clustering method we were able to capture it and identify its unique switching pattern that is repeated every time this event occurs. 

\begin{figure}[h!] 
\begin{center}
\includegraphics[scale=0.54]{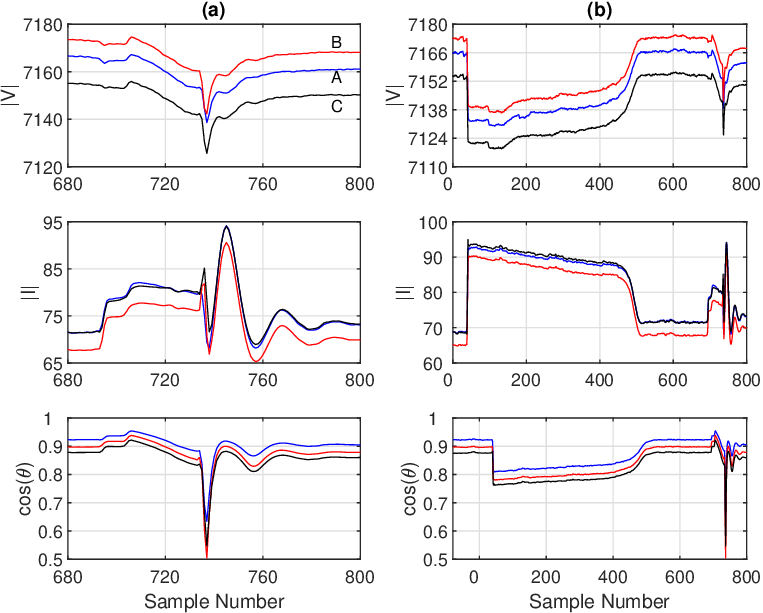}
\end{center}
\vspace{-0.2cm}
\caption{An example for the oscillation event in Cluster \#6: (a) oscillation in current; (b) the step change prior the oscillations.}
\label{fig:dynamic}
\vspace{-0.35cm}
\end{figure}

\subsubsection{Oscillations in Current Induced by Step Changes}
Fig. \ref{fig:dynamic} shows an example of an oscillation event in Cluster \#6. These events always occur immediately after a particular pattern of a step up change event in the current magnitude (as we can see at the beginning of the Fig. \ref{fig:dynamic}(a)) that also is followed by an oscillation event which is magnified in Fig. \ref{fig:dynamic}(b). \color{black}For  this  particular  class  of  oscillatory  events,  we  have  observed  that  the  median  for the  frequency  of  the  oscillations  is  5.17 Hz;  while  the  median  for  the  damping  ratio  of the  oscillations  is  2.64\%.  This  information  is  valuable  to  the  utility.  In  particular,  such information that is obtained in an unsupervised fashion by our proposed algorithms, when combined  with  a  subsequent  field  inspection  by  the  utility  crew  members,  can  quickly lead to the best remedial action; as deemed necessary by the utility. \color{black} This type of event causes the \emph{highest transient power factor} change at this distribution feeder, when compared with all kinds of events that we have captured in this study. The amount of the transient change in power factor is 0.4. 


 \vspace{0.3cm}

\subsection{Use Case Exposition: Three Clusters in Category \Rmnum{2}}

Three clusters are identified in Category \Rmnum{2}; denoted by Cluster \#7 to Cluster \#9. Clusters \#7 and \#8 can help identify voltage events. Cluster \#9 can identify voltage oscillations.

\subsubsection{Voltage Events} 

\begin{figure}[t] 
\begin{center}
\includegraphics[scale=0.54]{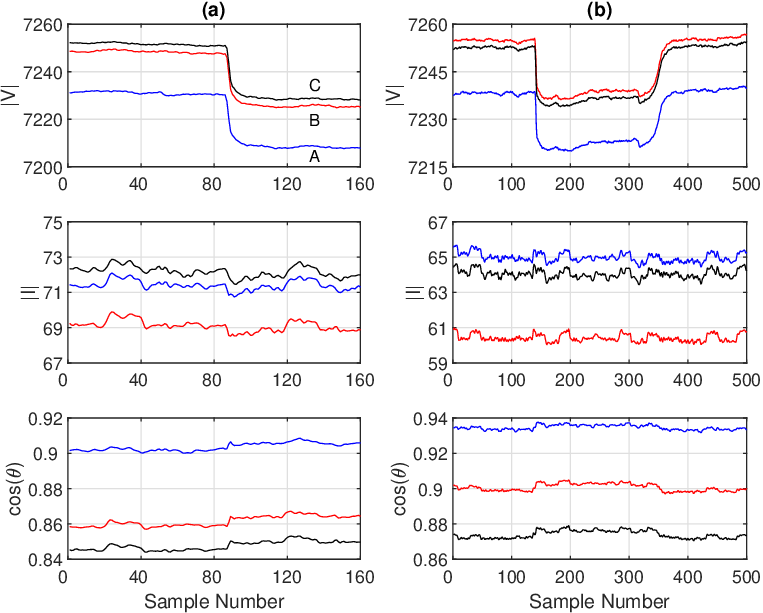}
\end{center}
\vspace{-0.25cm}
\caption{Examples of voltage events: (a) transformer tap-changer in Cluster \#7; (b) voltage  plateau in Cluster \#8.}
\vspace{-0.25cm}
\label{fig:vev}
\end{figure}

Fig. \ref{fig:vev}(a) shows an example of Cluster \#7, which is a transformer tap changing event. The events in this cluster inform the utility about voltage regulation status and the operation of tap-changers. Fig. \ref{fig:vev}(b) shows an example event in Cluster \#8, which is a voltage event with a \textit{plateau}.  The transient shape of the voltage in Cluster \#8 is similar to voltage changes in Cluster \#2, see Fig. \ref{fig:inrmed}(b); however, these two events are different because there is no change in current phasors ($\abs{I}$ and $cos(\theta)$) in the events in Cluster\#8. The events in Clusters \#7 and \#8 are often initiated at transmission level.

\begin{figure}[t] 
\begin{center}
\includegraphics[scale=0.54]{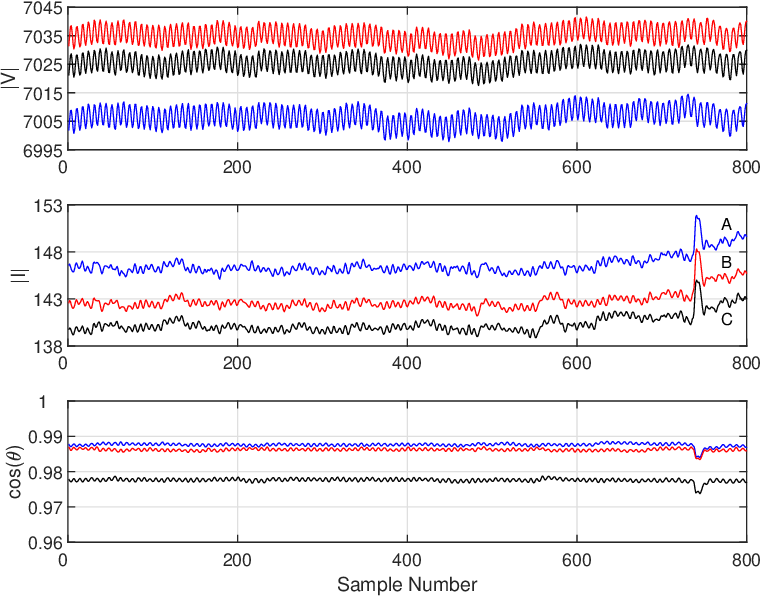}
\end{center}
\vspace{-0.25cm}
\caption{An example for voltage oscillation event in Cluster \#9.}
\vspace{-0.4cm}
\label{fig:vfreq}
\end{figure}

\begin{figure}[t] 
\begin{center}
\includegraphics[scale=0.54]{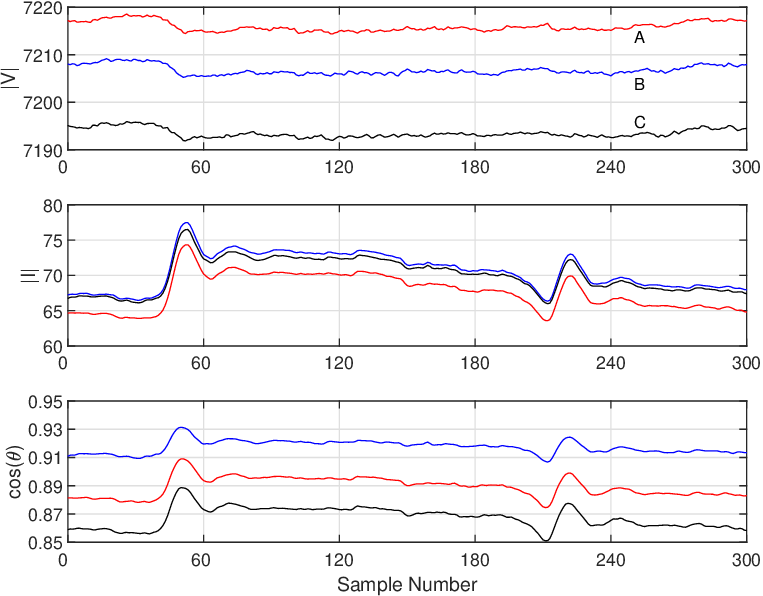}
\end{center}
\vspace{-0.25cm}
\caption{An example for current oscillation event in Cluster \#10.}
\label{fig:back}
\vspace{-0.1cm}
\end{figure}

\vspace{0.15cm}

\subsubsection{Voltage Oscillation Events} Fig. \ref{fig:vfreq} shows an example for an event in Cluster \#9, which is a high frequency low magnitude event in $\abs{V}$. Since there is no major change in current, this event can be due to two possible phenomena: 1) voltage oscillation from the upstream system; 2) temporary malfunction in micro-PMU data reporting. The later can be considered as a possibility if it persists and if other micro-PMUs do not report a similar behavior. In that case, this can be used as an indicator to request micro-PMU diagnostics. \color{black}Importantly, this cluster is a new cluster that is added by the active clustering method; which resulted from the significant difference between the samples in this cluster and the samples that were used during the offline training process.\color{black}

\vspace{0.3cm}

\subsection{Use Case Exposition: One Cluster in Category \Rmnum{3}}

One cluster is identified in Category \Rmnum{3}; denoted by Cluster \#10. Fig. \ref{fig:back} shows an example for this cluster. The events in this cluster affect only the current magnitude and power factor, rather than the voltage magnitude. It should be noted that, the pre-processing step in the proposed two-step clustering method helps to distinguish the events in Cluster \#10 from the events in Clusters \#2 and \#5, despite their relatively high MMC. 

\vspace{0.3cm}

\subsection{Use Case Exposition: Three Clusters in Category \Rmnum{4}}

\begin{figure}[t] 
\begin{center}
\includegraphics[scale=0.54]{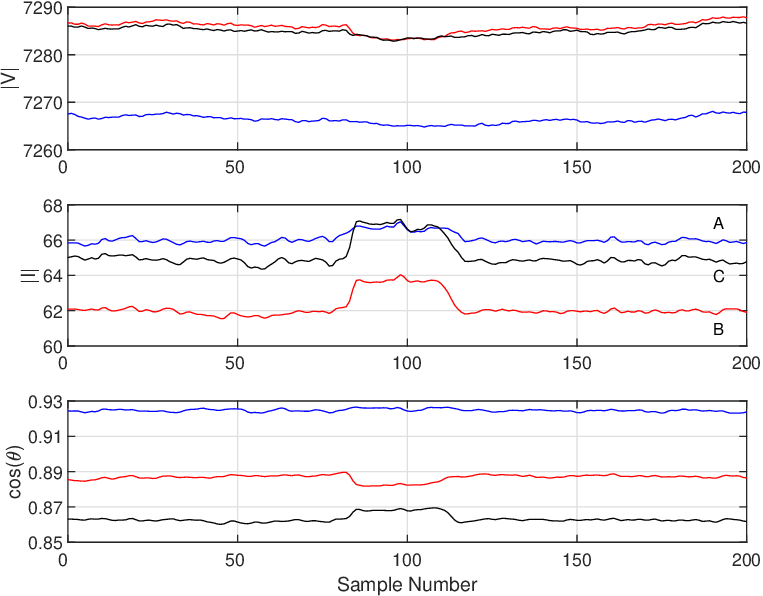}
\end{center}
\vspace{-0.25cm}
\caption{An example for the unbalanced events in Cluster \#11. The event affects the current magnitude of phases B and C.}
\label{fig:onetwo}
\vspace{-0.4cm}
\end{figure}

Three clusters are identified in Category \Rmnum{4}; denoted by Clusters \#11 to \#13. The events in these clusters are \emph{unbalanced}. Fig. \ref{fig:onetwo} shows an example of the event in Cluster \#11. This event is \emph{not} detected by the event detection method in \cite{arminevent}; because that method
fails to notice small changes in just one feature, i.e. in $\abs{I_{B}}$. However, in our method, by using one GAN model for each feature, even  small events are detected.

\vspace{0.3cm}

\subsection{Use Case Exposition: Three Clusters in Category \Rmnum{5}}
\begin{figure}[t] 
\begin{center}
\includegraphics[scale=0.54]{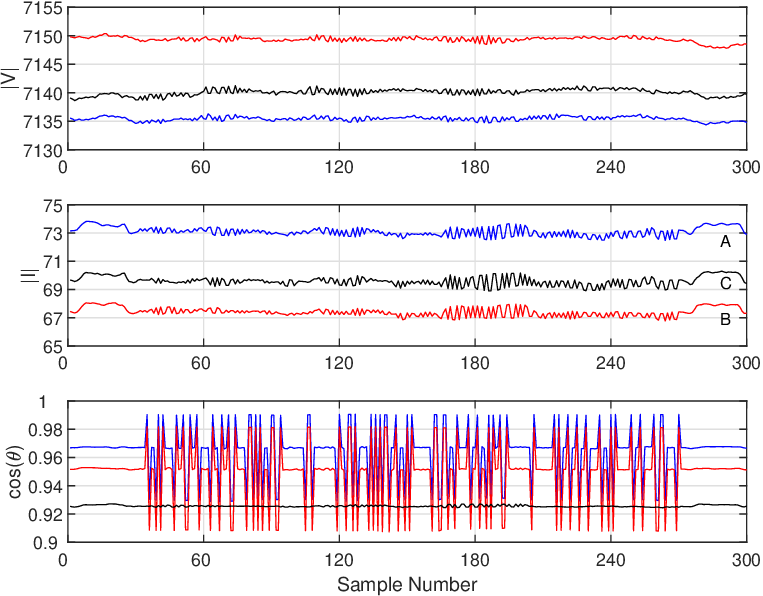}
\end{center}
\vspace{-0.25cm}
\caption{An example for power factor event in Cluster \#14.}
\vspace{-0.1cm}
\label{fig:noise}
\end{figure}

Three clusters are identified in Category \Rmnum{5}; denoted by Clusters \#14 to \#16. They are all related to power factor events. An example for an event in Cluster \#14 is shown in Fig. \ref{fig:noise}. It shows oscillations in power factor. There are also some minor oscillations, in the magnitudes of current and voltage during the same period. Other types of power factor events are also captured by the clusters in this category; not shown here. \color{black} It should be mentioned that the clusters in this category were added by the active clustering; i.e., they were not among the initial clusters that we had obtained during the offline training process. The creation of these new clusters was triggered mainly because of their different detection vectors.\color{black}

\vspace{0.3cm}
\subsection{Special Sequence of Events} \label{specific multiple event}

\begin{figure}[t] 
\begin{center}
\includegraphics[scale=0.54]{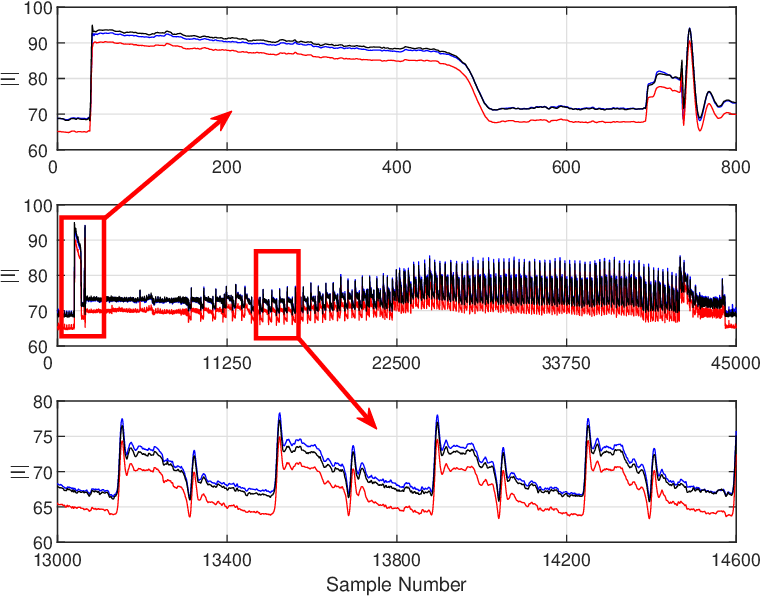}
\end{center}
\vspace{-0.25cm}
\caption{An example for the special sequence of the events in the current magnitude that are repeated occasionally. It was 
captured 
based on the collaboration of Clusters \#6 and \#10.}
\vspace{-0.4cm}
\label{fig:bigevent}
\end{figure}

One of the applications of the proposed unsupervised 
methods is to analyze the shape, occurrence time and sequence of the detected and clustered events. Our analysis shows that certain events come in sequence. 
This is an important observation to enhance the predictability of the system, its dynamics, and its events. An example is shown in 
Fig. \ref{fig:bigevent}. It is a \emph{super event} which consists of a sequence of several smaller events that belong to Clusters \#6 and \#10. This super event is first triggered by an event that belongs to Cluster \#6, which we previously saw in Fig. \ref{fig:dynamic}. Then, after about 60 seconds, a series of over 100 events occur that all belong to Cluster \#10. This sequence continues with a growing amplitude until it goes away. 
The exact same sequence of events occurred on the same day and around the same time each week.

\vspace{0.3cm}

\color{black}
\subsection{Versatility  of  the  Proposed  Model} \label{specific multiple event}
In order to show the versatility of the proposed event detection model, the developed model is applied to two other micro-PMU data set; which were not used to developed the existing model. First, a new data set that was from the \emph{neighboring} feeder is used for training but during a different time of the year. In this case, all we needed to do was to slightly fine tune the original model, i.e., we only needed to re-train the last layer in the existing GAN models, based on the new training batches, which lead to faster model training by using pre-trained model.  Thus, the training process in terms of computational time to achieve the equilibrium is around 14 minuets. It should be noted that, just like in the procedure mentioned in Section \ref{subsec: detection result}, a total of 1200 events were extracted manually within 6 hours. The MCC of the fine tuned model for this new data set is 0.9018.


Second, we used a micro-PMU data set from a \emph{completely different} type of feeder. This time we used real-world micro-PMU data from solar distribution feeder in a solar farm; based on the data in \cite{solar}. The nature of the power distribution feeder in this second case is drastically different from the nature of the original power distribution feeder that serves loads; which has been the focus throughout this paper. For the case of this second data set, we were able to keep the proposed architecture of our model; but we had to re-train the model with the new data set. It should be mentioned that the structure and the hyper-parameters (except for epoch and batch size numbers) remained the same as in our original model. Nevertheless, the result was promising. 
The results and other details about the analysis of the events at this solar farm are available in \cite{solar}.
\color{black}
\section{Conclusions}
\label{conslusion}

A set of new unsupervised methods are proposed to detect and cluster different types of events in micro-PMU measurements. 
The test results based on real-world micro-PMU data confirm that the proposed event detection method, which works based on training a novel GAN model, 
outperforms the existing, 
in particular when it comes to 
detecting the events 
that may impact only a subset of the features or only a subset of the phases
%
Test results also show the effectiveness of the proposed two-step clustering method, compared to the other prevalent methods, due to the proposed choice of the similarity measure and also the proposed architecture that improves clustering accuracy. 
Moreover, the active nature of the proposed clustering method makes it capable of identifying new clusters of events on \emph{an ongoing basis}. 
\color{black} Once the proposed unsupervised event detection and clustering are done, i.e., the events are detected and clustered, the results are
used in various used cases. Statistical analysis as well as human expert
knowledge are used to scrutinize the events in each cluster; to unmask
different applications for the utility operator. \color{black}


\bibliographystyle{IEEEtran}
\bibliography{main}

\end{document}